\documentclass[journal]{IEEEtran}

\usepackage{graphicx,latexsym,color}
\usepackage{amsmath,amsthm,amsfonts}
\usepackage{latexsym}
\usepackage{cite}

\definecolor{MyDarkBlue}{rgb}{0.1,0,0.55} 
\definecolor{MyRed}{rgb}{1,0,0} 

\usepackage{bm}
\usepackage{tikz}
\usepackage{pgfplots}
\usetikzlibrary{mindmap,trees}
\usetikzlibrary{%
  3d,
  calc,
  arrows,%
  shapes.misc,
  shapes.arrows,%
  chains,%
  matrix,%
  positioning,
  scopes,%
  decorations.pathmorphing,
  shadows%
}

\newcommand{\ie}{\textit{i.e.}\/, }
\newcommand{\eg}{\textit{e.g.}\/, }

\providecommand*{\mrm}[1]{\mathrm{#1}}
\providecommand*{\unit}[1]{\ensuremath{\mrm{\,#1}}}
\providecommand*{\eu}{\ensuremath{\mrm{e}}}
\providecommand*{\iu}{\ensuremath{\mrm{i}}}

\renewcommand{\Re}{\operatorname{Re}}	
\renewcommand{\Im}{\operatorname{Im}}	

\begin{document}

\title{Absorption and optimal plasmonic resonances for small ellipsoidal particles in lossy media}

\author{Mariana~Dalarsson, Sven~Nordebo,~\IEEEmembership{Senior Member,~IEEE}, Daniel~Sj\"{o}berg, Richard~Bayford
\thanks{Manuscript received \today. This work was supported by the Swedish Foundation for Strategic Research (SSF).}  
\thanks{Mariana~Dalarsson and Sven~Nordebo are with the Department of Physics and Electrical Engineering, Linn\ae us University, 
351 95 V\"{a}xj\"{o}, Sweden. E-mail: \{sven.nordebo,mariana.dalarsson\}@lnu.se.}
\thanks{Daniel~Sj\"{o}berg is with the Department of Electrical and Information Technology, Lund University, Box 118, 
221 00 Lund, Sweden. E-mail: daniel.sjoberg@eit.lth.se.}
\thanks{Richard~Bayford is with the Department of Natural Sciences, Middlesex University, Hendon campus, 
The Burroughs, London, NW4 4BT, United Kingdom. E-mail: R.Bayford@mdx.ac.uk.}
}


\maketitle

\begin{abstract}
A new simplified formula is derived for the absorption cross section of small dielectric 
ellipsoidal particles embedded in lossy media. 
The new expression leads directly to a closed form solution for the optimal conjugate match with respect to the surrounding medium, 
\ie the optimal permittivity of the ellipsoidal particle that maximizes the absorption at any given frequency. This defines the optimal plasmonic resonance for the ellipsoid.
The optimal conjugate match represents a metamaterial in the sense that the corresponding optimal permittivity function may have negative real part (inductive properties),
and can not in general be implemented as a passive material over a given bandwidth.
A necessary and sufficient condition is derived for the feasibility of tuning the Drude model to the optimal conjugate match at a single frequency, 
and it is found that all the prolate spheroids and some of the (not too flat) oblate spheroids can be tuned into optimal plasmonic resonance at any desired center frequency.
Numerical examples are given to illustrate the analysis. Except for the general understanding of  plasmonic resonances in lossy media,
it is  also anticipated that the new results can be useful for feasibility studies with \eg the radiotherapeutic hyperthermia based methods to treat cancer  
based on electrophoretic heating in gold nanoparticle suspensions using microwave radiation. 
\end{abstract}

\begin{IEEEkeywords}
Particle absorption, plasmonic resonances.
\end{IEEEkeywords}

\section{Introduction}\label{sect:introduction}
\IEEEPARstart{S}{}urface plasmon effects in gold nanoparticles is a physical phenomena that has been observed in colored glass objects since ancient times \cite{Maier2007}.
The most fascinating and useful features of the plasmonic resonances in metal nanoparticles is first of all the
mere existence of these resonances that may occur at free-space wavelengths that are many order of magnitudes
larger than the structure itself, and secondly (and contrary to intuition) that the corresponding resonance frequencies are virtually
independent of the size of the particles (if they are sufficiently small), but does depend on its shape and orientation, see \eg \cite{Link+etal2000,Maier2007}.
Today, new theory and applications of plasmonics are constantly being explored in technology, biology and medicin.
The topic includes the study of surface plasmonic resonances in small structures of various shapes, possibly embedded in different media,
see \eg \cite{Link+etal2000,Maier2007,Miller+etal2016}.
The present study is restricted to passive surrounding materials, but future applications of plasmonics may even include
amplifying (active) media as described in \eg \cite{Lawandy2004}.

The classical theories as well as most of the recent studies on plasmonic resonance effects are concerned with metal
nanoparticles and photonics where the exterior domain is lossless, see \eg \cite{Bohren+Huffman1983,Link+etal2000,Lawandy2004,Maier2007,Miller+etal2016}.
There are very few results developed for absorption and plasmonic resonance effects in particles or structures surrounded by lossy media.
As \eg in \cite{Miller+etal2016} is given geometry independent absorption bounds for the plasmonic resonances in metal nanoparticles in vacuum,
and an indication is given about how their results can be extended to lossy surrounding media.
There exists a general Mie theory for the electromagnetic power absorption in small spherical particles surrounded by lossy media, 
with explicit expressions and asymptotic formulas for the corresponding absorption cross section,
see \eg \cite{Chylek1977,Bohren+Gilra1979,Bohren+Huffman1983,Sassaroli+etal2012}. 
Even though these formulas are derived for spherical geometry, they are in general quite complicated and difficult to interpret. 
However, as will be demonstrated in this paper, a new simplified formula for the absorption cross section can be derived which is valid for small ellipsoidal particles
embedded in lossy media, and which facilitates a definition of the corresponding optimal plasmonic resonance.

A new potentially interesting application area for the plasmonic resonance phenomena is
with the electrophoretic heating of gold nanoparticle suspensions as a radiotherapeutic hyperthermia based 
method to treat cancer \cite{Lund+etal2011,Hanson+etal2011,Corr+etal2012,Sassaroli+etal2012,Collins+etal2014,Nordebo+etal2017a}. 
In particular, gold nanoparticles (GNPs) can be coated with ligands (nutrients) that target specific cancer cells as well as 
providing a net electronic charge of the GNPs \cite{Corr+etal2012,Sassaroli+etal2012,Collins+etal2014}. 
The hypothesis is that when a localized, charged GNP suspension has been taken up by the cancer cells, it will facilitate an 
electrophoretic current and a heating that can destroy the cancer under radio or microwave radiation,  
and this without causing damage to the normal surrounding tissues \cite{Corr+etal2012,Sassaroli+etal2012,Nordebo+etal2017a}.  
Hence, the potential medical application at  radio or microwave frequencies provides a motivation for studying optimal plasmonic resonances in
lossy media. However, it is also important to consider the complexity of this clinical application with many possible physical and 
biophysical phenomena to take into account, including cellular properties and their influence on the dielectric spectrum \cite{Gabriel+etal1996b,Lund+etal2011},
as well as thermodynamics and heat transfer, see \eg \cite{Hanson+etal2011}.
It is also interesting to note that several authors have questioned whether metal nanoparticles can be heated in radio frequency at all, see \eg \cite{Collins+etal2014,Hanson+etal2011}.
Based on the above mentioned results \cite{Hanson+etal2011,Corr+etal2012,Sassaroli+etal2012,Collins+etal2014,Nordebo+etal2017a}
as well as our own pre-studies in \cite{Nordebo+Sjoberg2016a}, we are proposing that straightforward physical modeling can be used to 
show that the most basic electromagnetic heating mechanisms, such as standard Joule heating and inductive heating, most likely can be disregarded 
for this medical application, whereas the potential application remains with the feasibility of achieving plasmonic (electrophoretic) resonances.

Recently, an optimal plasmonic resonance for the sphere has been defined as the optimal conjugate match with respect to the surrounding medium, 
\ie the optimal permittivity of the spherical suspension that maximizes the absorption at any given frequency \cite{Nordebo+etal2017a}.
It has been demonstrated in \cite{Nordebo+etal2017a} that for a surrounding medium consisting of a weak electrolyte solution (relevant for human tissue in the GHz range), 
a significant radio or microwave heating can be achieved inside a small spherical GNP suspension, provided that an electrophoretic particle acceleration (Drude) 
mechanism is valid and can be tuned into resonance at the desired frequency. 

In this paper, we generalize the results in \cite{Nordebo+etal2017a} to include small structures of ellipsoidal shapes embedded in lossy media, and we 
provide explicit expressions for the corresponding absorption cross section and optimal conjugate match (optimal plasmonic resonance). 
We investigate the necessary and sufficient condition regarding the feasibility of tuning a Drude model to optimal conjugate match at a single frequency,  
and we discuss the relation between the optimal conjugate match and the classical Fr\"{o}lich resonance condition.
A relative absorption ratio is defined to facilitate a quantitative and unitless indicator for the achievable local heating,
and some general expressions are finally given regarding the orientation of the ellipsoid in the polarizing field.
Numerical examples are included to illustrate the theory based on simple spheroidal geometries, and which at the same time are relevant 
for the potential medical application with electrophoretic heating
of GNP suspensions in the microwave regime.

\section{Optimal absorption in small ellipsoidal particles surrounded by lossy media}

\subsection{Notation and conventions}
The following notation and conventions will be used below.
Classical electromagnetic theory is considered based on SI-units \cite{Jackson1999}
and with time convention $\eu^{-\iu\omega t}$ for time harmonic fields. Hence, a passive dielectric material with relative permittivity $\epsilon$
has positive imaginary part.
Let $\mu_0$, $\epsilon_0$, $\eta_0$ and ${\rm c}_0$ denote the permeability, the permittivity, the wave impedance and
the speed of light in vacuum, respectively, and where $\eta_0=\sqrt{\mu_0/\epsilon_0}$ and ${\rm c}_0=1/\sqrt{\mu_0\epsilon_0}$.
The wavenumber of vacuum is given by $k_0=\omega\sqrt{\mu_0\epsilon_0}$, where $\omega=2\pi f$ is the angular frequency and $f$ the frequency.
The cartesian unit vectors are denoted $(\hat{\bm{x}}_1,\hat{\bm{x}}_2,\hat{\bm{x}}_3)$ and the radius vector is 
$\bm{r}=r\hat{\bm{r}}$ where $\hat{\bm{r}}$ is the radial unit vector in spherical coordinates.
Finally, the real and imaginary part and the complex conjugate of a complex number $\zeta$ are denoted 
$\Re\!\left\{\zeta\right\}$, $\Im\!\left\{\zeta\right\}$ and $\zeta^*$, respectively.

\subsection{Absorption and optimal resonances in spheres}
Consider a small spherical region of radius $r_1$ ($k_0r_1\ll 1$) consisting of a dielectric material with relative permittivity $\epsilon_1$ and which is suspended inside
a lossy dielectric background medium having relative permittivity $\epsilon$. Both media are assumed to be homogeneous and isotropic.
The absorption cross section $C_{\rm abs}$ of the small sphere is given by
\begin{eqnarray}\label{eq:Cabs}
C_{\rm abs}=C_{\rm ext}+C_{\rm amb}-C_{\rm sca},
\end{eqnarray}
where the scattering cross section $C_{\rm sca}$, the extinction cross section $C_{\rm ext}$ and 
the absorption cross section with respect to the ambient material $C_{\rm amb}$, are given by
\begin{eqnarray}
C_{\rm sca}=\frac{16\pi}{3}k_0r_1^3\Im\{\sqrt{\epsilon}\}\left|\frac{\epsilon_1-\epsilon}{\epsilon_1+2\epsilon} \right|^2, \label{eq:Csca} \\
C_{\rm ext}=6\pi k_0r_1^3\left[\frac{4}{9}\Re\left\{\frac{\epsilon_1-\epsilon}{\epsilon_1+2\epsilon}\right\}\Im\{\sqrt{\epsilon}\} \right. \nonumber \\
 \left. +\frac{2}{3}\Im\left\{\frac{\epsilon_1-\epsilon}{\epsilon_1+2\epsilon}\right\}
\left(\Re\{\sqrt{\epsilon}\}-\frac{(\Im\{\sqrt{\epsilon}\})^2}{\Re\{\sqrt{\epsilon}\}} \right) \right], \label{eq:Cext} \\
C_{\rm amb}=\frac{8\pi}{3}k_0r_1^3\Im\{\sqrt{\epsilon}\},\label{eq:Cinc}
\end{eqnarray}
see \eg \cite{Sassaroli+etal2012,Bohren+Huffman1983,Bohren+Gilra1979,Chylek1977}.
By algebraic manipulation of \eqref{eq:Cabs} through \eqref{eq:Cinc}, exploiting relations such as
$\Re\{\zeta\}=(\zeta+\zeta^*)/2$, $\Im\{\zeta\}=(\zeta-\zeta^*)/2\iu$ and $\Im\{\zeta\}=2\Re\{\sqrt{\zeta}\}\Im\{\sqrt{\zeta}\}$,
it can be shown that the absorption cross section can also be expressed in the simplified form
\begin{eqnarray}\label{eq:Cabs2}
C_{\rm abs}=12\pi k_0r_1^3 \frac{|\epsilon|^2}{\Re\{\sqrt{\epsilon}\}}\frac{\Im\{\epsilon_1\}}{\left|\epsilon_1+2\epsilon \right|^2},
\end{eqnarray}
see also \cite{Nordebo+etal2017a}. In particular, from \eqref{eq:Cabs2} it can be shown that the optimal conjugate match
$\epsilon_1^{\rm o}=-2\epsilon^*$ is the maximizer of  $C_{\rm abs}$ for $\Im\epsilon_1>0$, and which defines the
optimal plasmonic resonance for the sphere in a lossy surrounding medium \cite{Nordebo+etal2017a}.

The polarizability of the sphere is given by
\begin{equation}\label{eq:alphasphere}
\alpha=3V\frac{\epsilon_1-\epsilon}{\epsilon_1+2\epsilon},
\end{equation}
where $V=4\pi r_1^3/3$ is the volume of the spherical particle, see \eg \cite{Bohren+Huffman1983}. 
By inserting \eqref{eq:alphasphere} into \eqref{eq:Cabs} through \eqref{eq:Cinc}, the following expression can be obtained
\begin{multline}\label{eq:Cabs3}
C_{\rm abs}=k_0\Im\{\sqrt{\epsilon}\}\left(2V-\frac{4}{9V}|\alpha|^2+\frac{2}{3}\Re\{\alpha\} \right. \\
\left. -\frac{\Im\{\sqrt{\epsilon}\}}{\Re\{\sqrt{\epsilon}\}}\Im\{\alpha\} \right)
+k_0\Re\{\sqrt{\epsilon}\}\Im\{\alpha\}.
\end{multline}
Alternatively, \eqref{eq:alphasphere} can be rewritten as
\begin{equation}\label{eq:alphasphere2}
\frac{1}{\epsilon_1+2\epsilon}=\frac{\alpha}{3V(\epsilon_1-\epsilon)},
\end{equation}
and inserted into \eqref{eq:Cabs2} to yield
\begin{eqnarray}\label{eq:Cabs4}
C_{\rm abs}=\frac{k_0}{V}\frac{|\epsilon|^2}{\Re\{\sqrt{\epsilon}\}}\Im\{\epsilon_1\}\left|\frac{\alpha}{\epsilon_1-\epsilon}\right|^2.
\end{eqnarray}
At this point, it is emphasized that both expressions \eqref{eq:Cabs3} and \eqref{eq:Cabs4} have been derived based on the spherical assumption via \eqref{eq:alphasphere}.
When $\epsilon$ is real valued the expression \eqref{eq:Cabs3} reduces to the well known expression for the absorption cross section of small particles of arbitrary shape
that are surrounded by lossless media, \ie $C_{\rm abs}=C_{\rm ext}=k_0\sqrt{\epsilon}\Im\{\alpha\}$, see \cite{Bohren+Huffman1983}. 
On the other hand, the expression \eqref{eq:Cabs4} is in a more simple form which is well suited for the derivation of the optimal plasmonic resonance
in connection with \eqref{eq:alphasphere}.
It should be noted that the denominator $\epsilon_1-\epsilon$ in \eqref{eq:Cabs4} does not represent a pole of  
$C_{\rm abs}$ at $\epsilon_1=\epsilon$, its significance is instead to cancel the corresponding zero that is present in the polarizability $\alpha$ given by \eqref{eq:alphasphere}.

\subsection{Absorption in homogeneous ellipsoids}\label{sect:abshomellipsoids}
To derive the polarizability of a small homogeneous structure or a particle, it is assumed that the excitation is given by
a constant static electric field $\bm{E}_0=E_0\hat{\bm{x}}_j$, with the polarization defined by the direction of the $j$th cartesian axis.
The fundamental equations to be solved are given by
\begin{equation}\label{eq:fundamental}
\left\{\begin{array}{l}
\nabla\times\bm{E}(\bm{r})=\bm{0}, \vspace{0.2cm} \\
\nabla\cdot\bm{D}(\bm{r})=0, \vspace{0.2cm} \\
\bm{D}(\bm{r})=\epsilon_0\epsilon(\bm{r})\bm{E}(\bm{r}), \vspace{0.2cm} \\
\displaystyle \lim_{r\rightarrow\infty}\bm{E}(\bm{r})=\bm{E}_0,
\end{array}\right.
\end{equation}
where $\bm{E}(\bm{r})$ and $\bm{D}(\bm{r})$ are the electric field intensity and the electric flux density (electric displacement), respectively,
and where $\epsilon(\bm{r})$ denotes the complex valued relative permittivity which is assigned the appropriate constant values inside and outside the structure. 
The equations in \eqref{eq:fundamental} are solved by introducing the scalar potential $\Phi(\bm{r})$ where $\bm{E}(\bm{r})=-\nabla\Phi(\bm{r})$, 
and where $\Phi(\bm{r})$ satisfies the Laplace equation $\nabla^2\Phi(\bm{r})=0$, together with the continuity of $\Phi(\bm{r})$
as well as the continuity of the normal derivative $\epsilon(\bm{r})\frac{\partial}{\partial n}\Phi(\bm{r})$ at the boundary of the structure.
Finally, the scalar field must satisfy the asymptotic requirement $\lim_{r\rightarrow\infty}\Phi(\bm{r})=-E_0x_j$.
The resulting dipole moment relative the background is then given by
\begin{equation}\label{eq:pdef1}
\bm{p}=\int_{V}\epsilon_0(\epsilon_1-\epsilon)\bm{E}_1(\bm{r}){\rm d}v,
\end{equation}
where $\bm{E}_1(\bm{r})$ denotes the electric field inside the structure and the letter $V$ is used to denote
the domain of the structure as well as its volume.

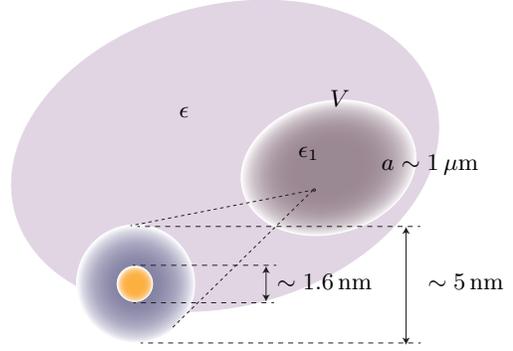
\begin{figure}[t]
\begin{picture}(50,150)
\put(40,10){\makebox(150,120){\includegraphics[width=6cm]{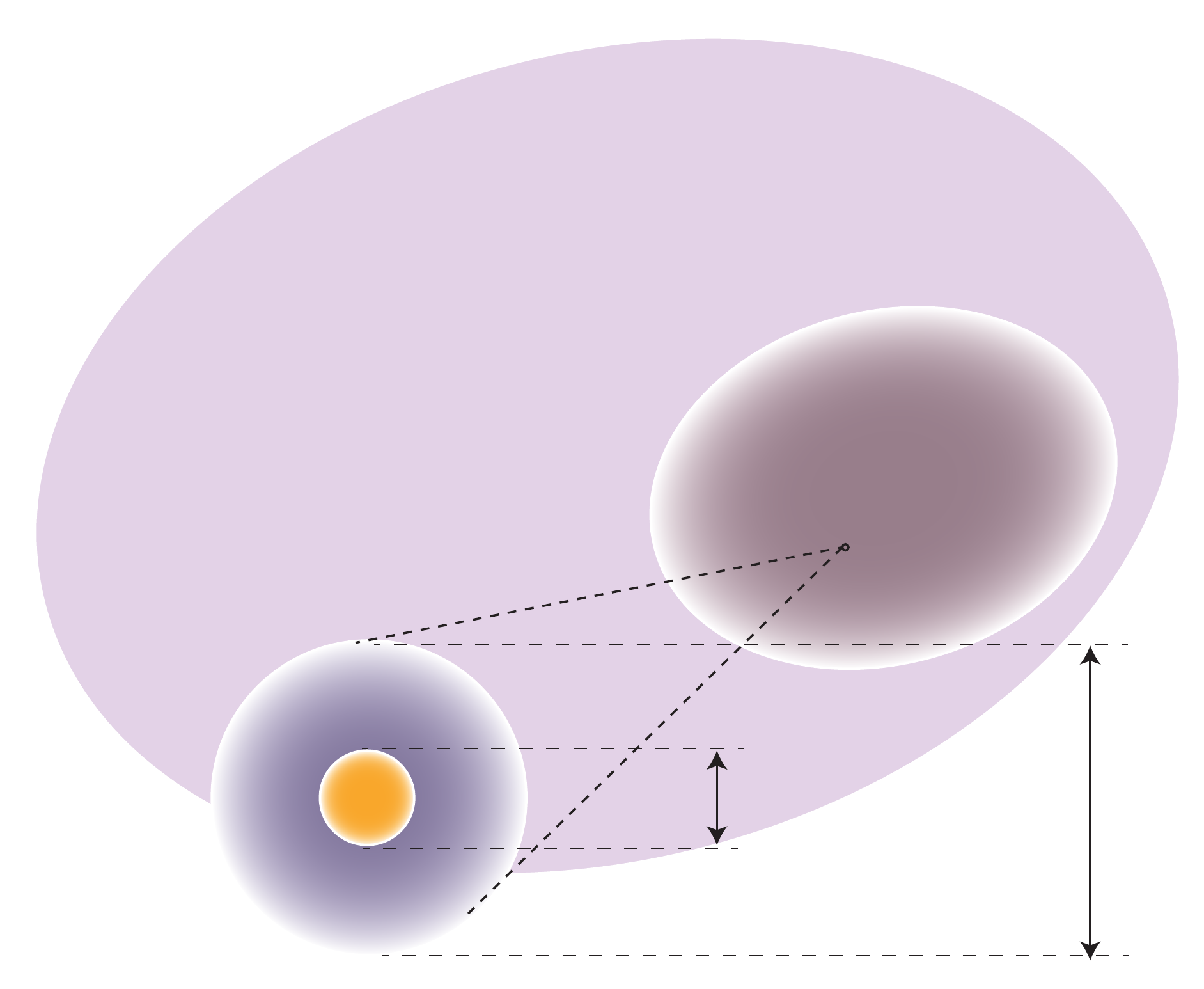}}} 
\put(152,94){ \small  $V$}
\put(140,75){ \small  $\epsilon_1$}
\put(95,90){ \small  $\epsilon$}
\put(132,25){ \small  $\sim 1.6$\unit{nm}}
\put(189,25){ \small  $\sim 5$\unit{nm}}
\put(175,70){\small $a\sim 1\unit{\mu m}$} 
\end{picture}
\caption{A small ellipsoidal region with permittivity $\epsilon_1$ and volume $V$, surrounded by a lossy
background material with permittivity $\epsilon$. The figure also illustrates some typical dimensions of
coated gold nanoparticles constituting the ellipsoidal suspension with spatial dimension $a$, see also \cite{marquez2013hyperthermia,Lund+etal2011}.}
\label{fig:RFsetup4pdf}
\end{figure}

Consider now a small ellipsoidal region consisting of a dielectric material with relative permittivity $\epsilon_1$ and volume $V$, 
and which is suspended inside a lossy dielectric background medium having relative permittivity $\epsilon$, see Figure \ref{fig:RFsetup4pdf}. 
Both media are assumed to be homogeneous and isotropic. Let the largest spatial dimension of the ellipsoid be denoted $a$ and assume that $k_0a\ll 1$. 
A solution to the electrostatic problem \eqref{eq:fundamental} for the ellipsoid is provided by \cite{Bohren+Huffman1983},
and it is shown that when the applied field is aligned along one of the axes of the ellipsoid the resulting
electric field $\bm{E}_1$ is constant inside the particle and parallel to the applied field $\bm{E}_0$.
From the analytical solution of this problem, the polarizability $\alpha_j$ of the ellipsoid is then finally obtained from the definition
\begin{equation}\label{eq:palphadef}
\bm{p}=\epsilon_0\epsilon\alpha_j\bm{E}_0.
\end{equation}
The resulting formula for the polarizability of the ellipsoid with 
semiaxes $a_i$ parallel to the cartesian axes $\hat{\bm{x}}_i$, $i=1,2,3$, and excitation $\bm{E}_0=E_0\hat{\bm{x}}_j$ is given by
\begin{equation}\label{eq:alphaellipsoid}
\alpha_j=3V\frac{\epsilon_1-\epsilon}{3\epsilon+3L_j(\epsilon_1-\epsilon)},
\end{equation}
where 
\begin{equation}
L_j=\frac{3V}{8\pi}\int_0^\infty\frac{{\rm d}q}{(a_j^2+q)f(q)},
\end{equation}
for $j=1,2,3$ and where $V=4\pi a_1a_2a_3/3$ and $f(q)=\sqrt{(q+a_1^2)(q+a_2^2)(q+a_3^2)}$, see \cite{Bohren+Huffman1983,Maier2007}. 
Here, $L_1$, $L_2$ and $L_3$ are geometrical factors satisfying $L_1+L_2+L_3=1$.

Note that  $\bm{p}$ is the additional dipole moment
added to the background polarization. This is obvious from the expression \eqref{eq:alphaellipsoid}
implying that $\bm{p}=\bm{0}$ when $\epsilon_1=\epsilon$. 
Note also that $\epsilon_1-\epsilon$ is the additional permittivity inside the particle with respect to the background. 
Hence, the total polarization of the medium inside the particle can be written
\begin{equation}
\bm{P}_1=\epsilon_0(\epsilon_1-1)\bm{E}_1=\epsilon_0(\epsilon-1)\bm{E}_1+\epsilon_0(\epsilon_1-\epsilon)\bm{E}_1,
\end{equation}
and the additional polarization $\bm{P}=\epsilon_0(\epsilon_1-\epsilon)\bm{E}_1$ yields the additional dipole moment relative the background
\begin{equation}\label{eq:pdef2}
\bm{p}=\int_{V}\epsilon_0(\epsilon_1-\epsilon)\bm{E}_1{\rm d}v=\epsilon_0(\epsilon_1-\epsilon)V\bm{E}_1,
\end{equation}
where $\bm{E}_1$ is a constant vector.
By comparison of \eqref{eq:palphadef} and \eqref{eq:pdef2}, and exploiting that $\bm{E}_0$ and $\bm{E}_1$ are parallel,
it is found that the interior field of the particle is given by
\begin{equation}\label{eq:E1sol}
\bm{E}_1=\frac{\epsilon\alpha_j}{V(\epsilon_1-\epsilon)}\bm{E}_0.
\end{equation}
The Poynting's theorem gives the total power loss inside the particle as
\begin{multline}\label{eq:Poyntingstheorem}
W_{\rm loss}=\frac{1}{2}\omega\epsilon_0\Im\{\epsilon_1\}\int_{V}\left|\bm{E}_1\right|^2{\rm d}v \\
=\frac{1}{2}k_0\eta_0^{-1}\Im\{\epsilon_1\}\frac{1}{V}\frac{|\epsilon|^2|\alpha_j|^2}{|\epsilon_1-\epsilon|^2}|E_0|^2,
\end{multline}
where \eqref{eq:E1sol} has been used.
The power density of a plane wave in a lossy medium is given by $P=\frac{1}{2}\Re \{E_0H_0^*\}$ where $H_0=E_0/\eta$ and $\eta=\eta_0/\sqrt{\epsilon}$.
Hence, the absorption cross section is finally obtained as
\begin{equation}\label{eq:Cabs5}
C_{\rm abs}=\frac{W_{\rm loss}}{|E_0|^2\frac{1}{2}\eta_0^{-1}\Re\{\sqrt{\epsilon}\}}=
\frac{k_0}{V}\frac{|\epsilon|^2}{\Re\{\sqrt{\epsilon}\}}\Im\{\epsilon_1\}\left|\frac{\alpha_j}{\epsilon_1-\epsilon}\right|^2,
\end{equation}
which is identical to the formula given in \eqref{eq:Cabs4}.

\subsection{Optimal plasmonic resonances for the ellipsoid}\label{sect:optplasmresellipsoid}
Consider the real valued function 
\begin{equation}\label{eq:F0def}
g(\epsilon_1)=\frac{\Im\epsilon_1}{|\epsilon_1-{\epsilon_1^{\rm o}}^*|^2},
\end{equation}
where $\epsilon_1$ is a complex variable with $\Im\{\epsilon_1\}> 0$ and $\epsilon_1^{\rm o}$ a constant with $\Im\{\epsilon_1^{\rm o}\}> 0$.
Take the complex derivative of $g(\epsilon_1)$ with respect to $\epsilon_1^*$ to yield
\begin{equation}
\frac{\partial}{\partial\epsilon_1^*}g(\epsilon_1)=\frac{\iu}{2}\frac{1}{|\epsilon_1-{\epsilon_1^{\rm o}}^*|^2}\frac{\epsilon_1-\epsilon_{1}^{\rm o}}{\epsilon_1^*-\epsilon_{1}^{\rm o}},
\end{equation}
showing that $\epsilon_1=\epsilon_{1}^{\rm o}$ is a stationary point.
It has furthermore been shown in \cite{Nordebo+etal2017a} that $g(\epsilon_1)$ is a strictly concave function with a local maximum at $\epsilon_1=\epsilon_1^{\rm o}$,
and hence we refer to $\epsilon_1^{\rm o}$ as the optimal conjugate match. 

The absorption cross section \eqref{eq:Cabs5} for the ellipsoid with polarizability \eqref{eq:alphaellipsoid}, is given by
\begin{equation}\label{eq:Cabsellipsoid}
C_{\rm abs}=k_0V \frac{|\epsilon|^2}{\Re\{\sqrt{\epsilon}\}}\frac{1}{L_j^2}\frac{\Im\{\epsilon_1\}}{\left|\epsilon_1-\epsilon\frac{L_j-1}{L_j}\right|^2}.
\end{equation}
By comparison of \eqref{eq:F0def} and \eqref{eq:Cabsellipsoid}, it is immediately seen that the optimal conjugate match for the ellipsoid is given by
\begin{equation}\label{eq:epsilon1odef}
\epsilon_1^{\rm o}=-\epsilon^*\frac{1-L_j}{L_j},
\end{equation}
and which hence defines the optimal plasmonic resonance for the ellipsoid in a lossy surrounding medium.
The sphere is a special case of the ellipsoid with $L_1=L_2=L_3=1/3$ yielding $\epsilon_1^{\rm o}=-2\epsilon^*$,
and which reproduces the corresponding result given in \cite{Nordebo+etal2017a}.

The notion of the optimal resonance defined in \eqref{eq:epsilon1odef} as being ``plasmonic'' is motivated by the fact that a ``normal'' lossy background 
medium would have $\Re\{\epsilon\}>0$ and hence $\Re\{\epsilon_1^{\rm o}\}<0$, which is a typical feature of plasmonic resonances
and which can be achieved \eg by tuning a Drude model. If we consider the optimal conjugate match $\epsilon_1^{\rm o}$ in \eqref{eq:epsilon1odef} as a function of frequency, 
then it represents a metamaterial in the sense that it has a negative real part (a dielectric medium with inductive properties), and which can not in general be implemented as a passive material over a fixed bandwidth, see also \cite{Gustafsson+Sjoberg2010a,Nordebo+etal2017a}. However, as will be shown below, in many cases a Drude model can be tuned to optimal plasmonic resonance at any desired center frequency.

\subsection{Tuning the Drude model for the ellipsoid in a lossy surrounding medium}
A generalized Drude model for the permittivity of the ellipsoidal particle is given by 
\begin{eqnarray}\label{eq:epsilon1def}
\epsilon_1(\omega)=\epsilon(\omega)+\iu\frac{\sigma_1}{\omega\epsilon_0}\frac{1}{1-\iu\omega\tau_1},
\end{eqnarray}
where $\epsilon(\omega)$ corresponds to the background material and where the static conductivity $\sigma_1$ and the relaxation
time $\tau_1$ are the parameters of the additional Drude model. It is assumed that the background material is a ``normal'' material with $\Re\{\epsilon(\omega)\}>0$
and $\Im\{\epsilon(\omega)\}>0$ over the bandwidth of interest. The Drude parameters may correspond to \eg an electrophoretic mechanism where
$\sigma_1={\cal N}q^2/\beta$ and $\tau_1=m/\beta$, where ${\cal N}$ is the number of charged particles per unit volume, 
$q$ the particle charge, $\beta$ the friction constant of the host medium and $m$ the mass of the particle, see \eg \cite{Sassaroli+etal2012}.
The Drude parameters can be tuned to the optimal conjugate match  by solving the equation
\begin{eqnarray}\label{eq:tuningDrude1}
\epsilon(\omega_{\rm d})+\iu\frac{\sigma_1}{\omega_{\rm d}\epsilon_0}\frac{1}{1-\iu\omega_{\rm d}\tau_1}=\epsilon_1^{\rm o}(\omega_{\rm d}),
\end{eqnarray}
where  $\epsilon_1^{\rm o}(\omega_{\rm d})$ is given by \eqref{eq:epsilon1odef} and $\omega_{\rm d}=2\pi f_{\rm d}>0$ 
is the desired resonance frequency. 
This means that the following two equations corresponding to the real and imaginary parts of \eqref{eq:tuningDrude1} must be satisfied
\begin{equation}\label{eq:tuningDrude2}
\left\{\begin{array}{l}
\displaystyle \frac{\sigma_1\tau_1}{\epsilon_0(1+\omega_{\rm d}^2\tau_1^2)}=\Re\{\epsilon\}-\Re\{\epsilon_1^{\rm o}\}, \vspace{0.2cm} \\
\displaystyle \frac{\sigma_1}{\epsilon_0\omega_{\rm d}(1+\omega_{\rm d}^2\tau_1^2)}=\Im\{\epsilon_1^{\rm o}\}-\Im\{\epsilon\}.
\end{array}\right.
\end{equation}
To find a solution to \eqref{eq:tuningDrude2}, it is necessary and sufficient that both equations have a right-hand side that is positive.
For a ``normal'' surrounding material with $\Re\{\epsilon\}>0$, it is readily seen from \eqref{eq:epsilon1odef} that $\Re\{\epsilon_1^{\rm o}\}<0$
and hence that $\Re\{\epsilon\}-\Re\{\epsilon_1^{\rm o}\}>0$.
For the imaginary part, the requirement that $\Im\{\epsilon_1^{\rm o}\}-\Im\{\epsilon\}> 0$ together with \eqref{eq:epsilon1odef} leads directly to the condition
\begin{equation}\label{eq:Ljcond}
L_j<\frac{1}{2}.
\end{equation}
When the condition \eqref{eq:Ljcond} is fulfilled, the system \eqref{eq:tuningDrude2} can be solved to yield the following tuned Drude parameters
\begin{equation}\label{eq:tuningDrude3}
\left\{\begin{array}{l}
\displaystyle \tau_1=\frac{1}{\omega_{\rm d}}\frac{\Re\epsilon(\omega_{\rm d})-\Re\epsilon_1^{\rm o}(\omega_{\rm d})}
{\Im\epsilon_1^{\rm o}(\omega_{\rm d})-\Im\epsilon(\omega_{\rm d})}, \vspace{0.2cm} \\
\displaystyle \sigma_1=\epsilon_0\left(\Re\epsilon(\omega_{\rm d})-\Re\epsilon_1^{\rm o}(\omega_{\rm d}) \right)\frac{1+\omega_{\rm d}^2\tau_1^2}{\tau_1},
\end{array}\right.
\end{equation}
see also \cite{Nordebo+etal2017a}.

Consider the interpretation of the condition \eqref{eq:Ljcond} in the case with spheroidal shapes. 
Choose for example the $a_3$ axis as the direction of the applied electric field $\bm{E}_0=E_0\hat{\bm{x}}_3$, and let $L_3=1-2L$ where $L=L_1=L_2$. 
The ellipsoid is then a prolate spheroid when $L_3<1/3$, a sphere when $L_3=1/3$ and an oblate spheroid when $L_3>1/3$. The interpretation
of \eqref{eq:Ljcond} is that the sphere and the prolate spheriod can always be tuned by a Drude model to match
the optimal value $\epsilon_1^{\rm o}(\omega_{\rm d})$ at any desired center frequency $\omega_{\rm d}$ for which $\Re\epsilon(\omega_{\rm d})>0$. An oblate spheroid, however, 
can only be tuned into optimal plasmonic resonance using the Drude model \eqref{eq:epsilon1def}, when the shape is not too flat and $L_3<\frac{1}{2}$.
This result agrees well with intuition, since polarizability (and hence resonance) is enhanced by prolongation of the particle shape in the direction of the polarizing field.

\subsection{Relation to the Fr\"{o}hlich condition}
The result \eqref{eq:epsilon1odef} generalizes the classical Fr\"{o}hlich condition \cite{Maier2007} in the sense that \eqref{eq:epsilon1odef}
gives the condition for an optimal plasmonic resonance of a small homogeneous ellipsoid, which is not an approximation and which is valid for a surrounding lossy medium.
Hence, the Fr\"{o}hlich condition for the ellipsoid can be obtained from \eqref{eq:epsilon1odef} in a sequence of approximations as follows.
First, the criterion \eqref{eq:epsilon1odef} is approximated as 
\begin{equation}\label{eq:Frohlich1}
\Re\{\epsilon_1\}=\frac{L_j-1}{L_j}\Re\{\epsilon\},
\end{equation}
assuming that the imaginary parts of both $\epsilon$ and $\epsilon_1$ are small.
Using the following form of the Drude model
\begin{equation}\label{eq:FrohlichDrude}
\epsilon_1(\omega)=\epsilon(\omega)-\frac{\omega_{\rm p}^2\tau_1^2}{1+\omega^2\tau_1^2}+\iu\frac{\omega_{\rm p}^2\tau_1}{\omega(1+\omega^2\tau_1^2)},
\end{equation}
where $\omega_{\rm p}$ is the plasma frequency given by $\omega_{\rm p}^2=\sigma_1/(\epsilon_0\tau_1)$,
the equation \eqref{eq:Frohlich1} can be solved to yield the following Fr\"{o}hlich resonance frequency
\begin{equation}
\omega_0=\sqrt{\omega_{\rm p}^2\frac{L_j}{\Re\{\epsilon\}}-\frac{1}{\tau_1^2}}\approx\omega_{\rm p}\sqrt{\frac{L_j}{\Re\{\epsilon\}}},
\end{equation}
where the last approximation is valid when $\omega_0\tau_1\gg 1$.
For a lossless surrounding medium with real valued $\epsilon$, the Fr\"{o}hlich resonance frequency for a sphere consisting of a Drude metal
is given by $\omega_0=\omega_{\rm p}/\sqrt{3\epsilon}$, see \cite{Maier2007}.

\subsection{Relative absorption ratio}
The absorption cross section of a small volume with respect to the ambient material is given by $C_{\rm amb}=2k_0V\Im\{\sqrt{\epsilon}\}$, 
and which is valid for volumes of arbitrary shape, see also \eqref{eq:Cinc}.
A unitless relative absorption ratio for the ellipsoid can now be defined as
\begin{equation}\label{eq:Fdef}
F_{\rm abs}=\frac{C_{\rm abs}}{C_{\rm amb}}=\frac{|\epsilon|^2}{\Im\{\epsilon\}}\frac{1}{L_j^2}\frac{\Im\{\epsilon_1\}}{\left|\epsilon_1-\epsilon\frac{L_j-1}{L_j}\right|^2},
\end{equation}
where \eqref{eq:Cabsellipsoid} has been used, as well as the relationship $\Im\{\epsilon\}=2\Re\{\sqrt{\epsilon}\}\Im\{\sqrt{\epsilon}\}$.
By inserting the optimal conjugate match \eqref{eq:epsilon1odef} into \eqref{eq:Fdef}, the following optimal relative absorption ratio
is obtained for excitation along the $a_j$ axis of the ellipsoid
\begin{equation}\label{eq:Fmaxdef}
F_{\rm abs}^{\rm o}=\frac{|\epsilon|^2}{4(\Im\{\epsilon\})^2}\frac{1}{L_j(1-L_j)}.
\end{equation}

The relative absorption ratio given by \eqref{eq:Fdef} and \eqref{eq:Fmaxdef}
can be useful as a quantitative unitless measure showing how
much more heating that potentially can be obtained in a small resonant region in comparison to the ambient local heating.
It is important to note, however, that a complete system analysis would take into account not only the local heating capabilities,
but also the significance of the frequency dependent penetration (skin) depth of the bulk material, see also \cite{Nordebo+etal2017a}.

\subsection{General polarization}
Finally, a general expression is given for the absorption cross section of a small homogeneous ellipsoidal particle with arbitrary orientation with respect to the applied field.
Consider a small ellipsoidal region with its semiaxes $a_i$ aligned along the cartesian unit vectors $\hat{\bm{x}}_i$, $i=1,2,3$,
and an applied electric field given by $\bm{E}_0=E_{01}\hat{\bm{x}}_1+E_{02}\hat{\bm{x}}_2+E_{03}\hat{\bm{x}}_3$.
Due to the linearity of the fundamental equations \eqref{eq:fundamental}, it is straightforward to generalize the expressions on absorption cross section given in sections
\ref{sect:abshomellipsoids} and \ref{sect:optplasmresellipsoid} above.
The polarizability \eqref{eq:alphaellipsoid} can now be expressed in terms of the diagonal
polarizability dyadic $\bm{\alpha}=\alpha_1\hat{\bm{x}}_1\hat{\bm{x}}_1+\alpha_2\hat{\bm{x}}_2\hat{\bm{x}}_2+\alpha_3\hat{\bm{x}}_3\hat{\bm{x}}_3$
where $\bm{p}=\epsilon_0\epsilon\bm{\alpha}\cdot\bm{E}_0$,
and the interior field $\bm{E}_1$  is given by
\begin{equation}\label{eq:E1solgeneralized}
\bm{E}_1=\frac{\epsilon}{V(\epsilon_1-\epsilon)}\bm{\alpha}\cdot\bm{E}_0,
\end{equation}
instead of \eqref{eq:E1sol}. The total power loss inside the particle is now given by
\begin{multline}\label{eq:Poyntingstheoremgeneralized}
W_{\rm loss}=\frac{1}{2}\omega\epsilon_0\Im\{\epsilon_1\}\int_{V}\left|\bm{E}_1\right|^2{\rm d}v \\
=\frac{1}{2}k_0\eta_0^{-1}\Im\{\epsilon_1\}\frac{|\epsilon|^2}{V|\epsilon_1-\epsilon|^2}\sum_{j=1}^3|\alpha_j|^2|E_{0j}|^2,
\end{multline}
and the corresponding absorption cross section 
\begin{multline}\label{eq:Cabs5generalized}
C_{\rm abs}=\frac{W_{\rm loss}}{|E_0|^2\frac{1}{2}\eta_0^{-1}\Re\{\sqrt{\epsilon}\}} \\
=\frac{k_0}{V}\frac{|\epsilon|^2}{\Re\{\sqrt{\epsilon}\}}\Im\{\epsilon_1\}\sum_{j=1}^3\frac{|E_{0j}|^2}{|E_0|^2}\left|\frac{\alpha_j}{\epsilon_1-\epsilon}\right|^2,
\end{multline}
where $|E_0|^2=|E_{01}|^2+|E_{02}|^2+|E_{03}|^2$.
By using \eqref{eq:alphaellipsoid}, the absorption cross section and the relative absorption ratio finally becomes
\begin{equation}\label{eq:Cabsellipsoidgeneralized}
C_{\rm abs}=k_0V \frac{|\epsilon|^2}{\Re\{\sqrt{\epsilon}\}}\sum_{j=1}^3\frac{|E_{0j}|^2}{|E_0|^2L_j^2}\frac{\Im\{\epsilon_1\}}{\left|\epsilon_1-\epsilon\frac{L_j-1}{L_j}\right|^2},
\end{equation}
and
\begin{equation}\label{eq:Fdefgeneralized}
F_{\rm abs}=\frac{C_{\rm abs}}{C_{\rm amb}}=\frac{|\epsilon|^2}{\Im\{\epsilon\}}\sum_{j=1}^3\frac{|E_{0j}|^2}{|E_0|^2L_j^2}\frac{\Im\{\epsilon_1\}}{\left|\epsilon_1-\epsilon\frac{L_j-1}{L_j}\right|^2}.
\end{equation}
It is immediately seen that the two expressions in \eqref{eq:Cabsellipsoidgeneralized} and \eqref{eq:Fdefgeneralized} are strictly concave
functions in terms of the complex variable $\epsilon_1$ for $\Im\{\epsilon_1\}> 0$ (a positive combination of concave functions is a concave function, etc)
and the corresponding optimal plasmonic resonance is therefore well-defined and unique. However, it is no longer possible to obtain a simple
closed form expression for the optimal conjugate match $\epsilon_{1}^{\rm o}$ as in \eqref{eq:epsilon1odef}.

\section{Numerical examples}
To illustrate the theory, a numerical example is considered with parameter choices relevant for the application
with microwave absorption in gold nanoparticle suspensions, see \eg \cite{Sassaroli+etal2012,Collins+etal2014,Nordebo+etal2017a}.
Hence, the resonant frequency is chosen here as $f_{\rm d}=1\unit{GHz}$ to mimic a typical system operating in the microwave regime,
see \eg \cite{marquez2013hyperthermia}. As a lossy ambient medium is taken the typical characteristics of human tissue.
Information about the dielectric properties of biological tissues can be found in \eg
\cite{Gabriel+etal1996b} giving measurement results of most organs including brain (grey matter), heart muscle, kidney, liver, inflated lung, spleen, muscle, etc.
From these measurement results we conclude that human tissue can be realistically modelled by using a conductivity
in the order of $1$\unit{S/m} and a permittivity similar to water at a frequency of $1$\unit{GHz}.
Hence, a simple conductivity-Debye model for saline water is considered here
where the surrounding medium is a weak electrolyte solution with relative permittivity 
\begin{eqnarray}\label{eq:epsilondef}
\epsilon(\omega)=\epsilon_{\infty}+\frac{\epsilon_{\rm s}-\epsilon_{\infty}}{1-\iu\omega\tau}+\iu\frac{\sigma}{\omega\epsilon_0},
\end{eqnarray}
where $\epsilon_{\infty}$, $\epsilon_{\rm s}$ and $\tau$ are the high frequency permittivity,  the static permittivity and the dipole relaxation time in the 
corresponding Debye model for water, respectively, and $\sigma$ the conductivity of the saline solution. 
In the numerical examples below, these parameters are chosen as $\epsilon_{\infty}=5.27$, $\epsilon_{\rm s}=80$, $\tau=10^{-11}$\unit{s} and 
$\sigma\in\{0.1,1\}$\unit{S/m}. 

\begin{figure}[htb]
\begin{picture}(50,140)
\put(100,0){\makebox(50,130){\includegraphics[width=8.5cm]{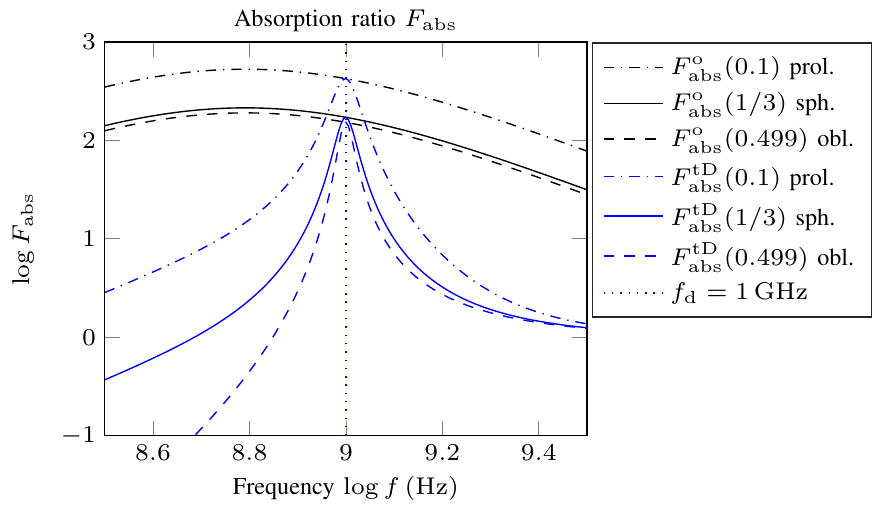}}} 
\end{picture}
\caption{Optimal and tuned absorption ratios $F_{\rm abs}$ with $F_{\rm abs}^{\rm o}(L_3)$ corresponding to the
optimal parameter $\epsilon_{1}^{\rm o}$ and $F_{\rm abs}^{\rm tD}(L_3)$ corresponding to the
tuned Drude parameter $\epsilon_{\rm 1}^{\rm tD}$. 
Here, the surrounding medium is a saline solution with $\sigma=0.1$\unit{S/m}.}
\label{fig:matfig503}
\end{figure}

\begin{figure}[htb]
\begin{picture}(50,140)
\put(100,0){\makebox(50,130){\includegraphics[width=8.5cm]{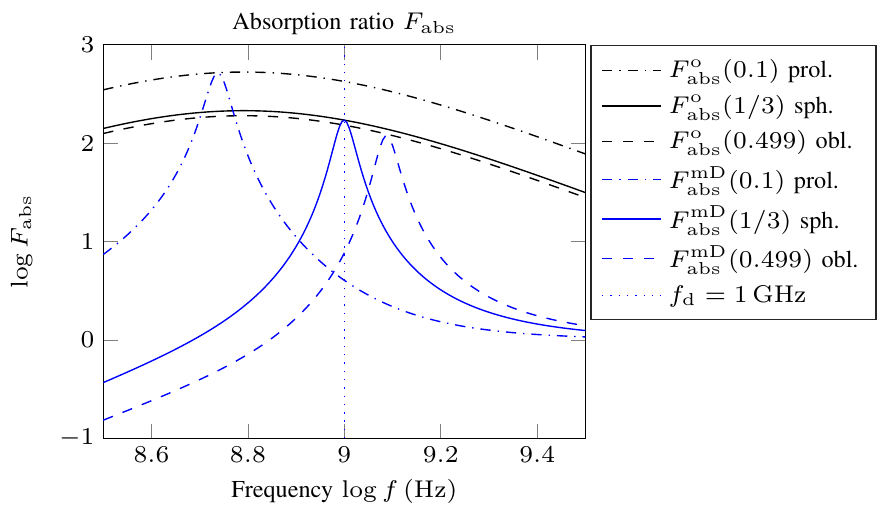}}}  
\end{picture}
\caption{Optimal and mismatched absorption ratios $F_{\rm abs}$ with $F_{\rm abs}^{\rm o}(L_3)$ corresponding to the
optimal parameter $\epsilon_{1}^{\rm o}$ and $F_{\rm abs}^{\rm mD}(L_3)$ corresponding to the mismatched
Drude parameter $\epsilon_{\rm 1}^{\rm mD}$ tuned to a sphere.
Here, the surrounding medium is a saline solution with $\sigma=0.1$\unit{S/m}.}
\label{fig:matfig504}
\end{figure}

\begin{figure}[htb]
\begin{picture}(50,140)
\put(100,0){\makebox(50,130){\includegraphics[width=8.5cm]{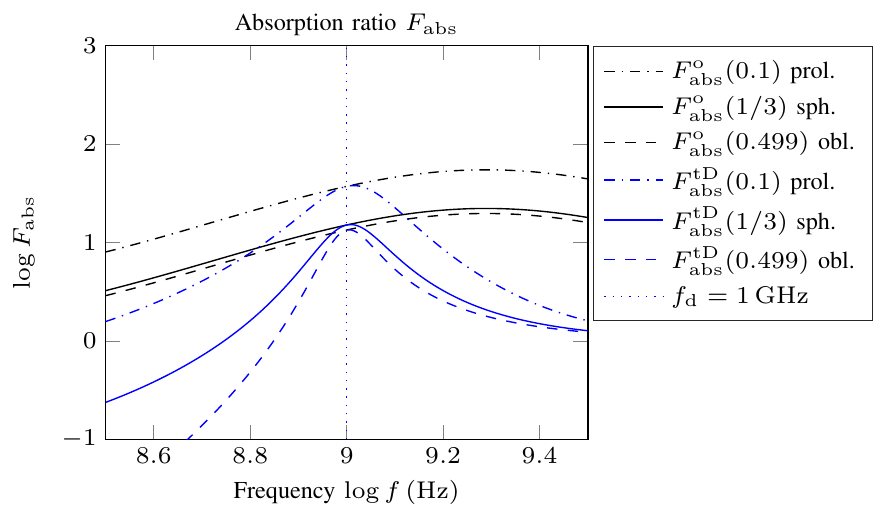}}} 
\end{picture}
\caption{Optimal and tuned absorption ratios $F_{\rm abs}$ with $F_{\rm abs}^{\rm o}(L_3)$ corresponding to the
optimal parameter $\epsilon_{1}^{\rm o}$ and $F_{\rm abs}^{\rm tD}(L_3)$ corresponding to the
tuned Drude parameter $\epsilon_{\rm 1}^{\rm tD}$. 
Here, the surrounding medium is a saline solution with $\sigma=1$\unit{S/m}.}
\label{fig:matfig503b}
\end{figure}

\begin{figure}[htb]
\begin{picture}(50,140)
\put(100,0){\makebox(50,130){\includegraphics[width=8.5cm]{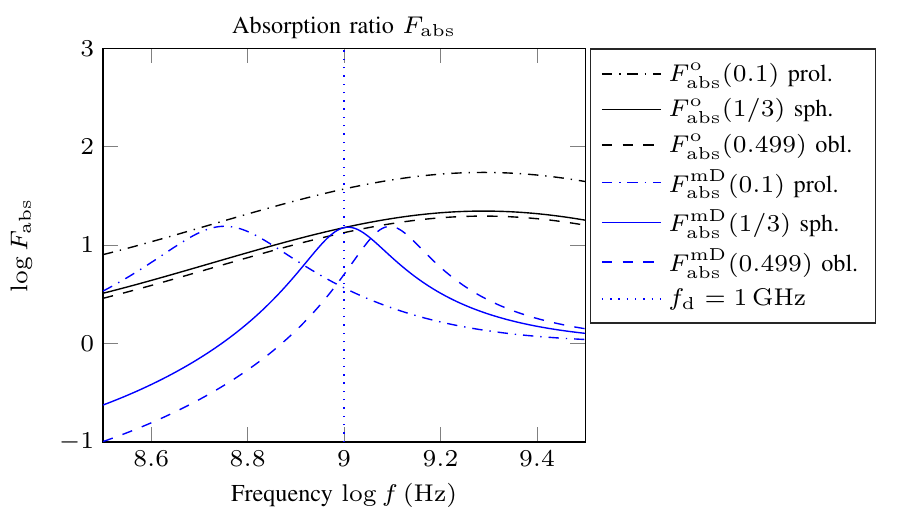}}} 
\end{picture}
\caption{Optimal and mismatched absorption ratios $F_{\rm abs}$ with $F_{\rm abs}^{\rm o}(L_3)$ corresponding to the
optimal parameter $\epsilon_{1}^{\rm o}$ and $F_{\rm abs}^{\rm mD}(L_3)$ corresponding to the mismatched
Drude parameter $\epsilon_{\rm 1}^{\rm mD}$ tuned to a sphere.
Here, the surrounding medium is a saline solution with $\sigma=1$\unit{S/m}.}
\label{fig:matfig504b}
\end{figure}

In Figures \ref{fig:matfig503} through \ref{fig:matfig504b} are shown the calculated relative absorption ratios \eqref{eq:Fdef}
for the ellipsoid with optimal, tuned Drude and mismatched Drude parameters, respectively.
The optimal parameter $\epsilon_1^{\rm o}$ is given by \eqref{eq:epsilon1odef}, the tuned Drude parameter $\epsilon_{\rm 1}^{\rm tD}$
is given by \eqref{eq:epsilon1def} and \eqref{eq:tuningDrude3}, and the mismatched Drude parameter $\epsilon_{\rm 1}^{\rm mD}$
is again the Drude parameter given by \eqref{eq:epsilon1def} and \eqref{eq:tuningDrude3}, but which is constantly mismatched to the sphere using 
$\epsilon_1^{\rm o}=-2\epsilon^*$.
A spheroidal shape is considered with the geometrical factors $L_3=1-2L$ and $L=L_1=L_2$, and where the applied electric field $\bm{E}_0=E_0\hat{\bm{x}}_3$
is aligned along the $a_3$ axis of the spheroid and hence $\alpha_3$ is given by \eqref{eq:alphaellipsoid}. The relative absorption ratios \eqref{eq:Fdef}
in the three cases described above are denoted 
$F_{\rm abs}^{\rm o}(L_3)$, $F_{\rm abs}^{\rm tD}(L_3)$ and $F_{\rm abs}^{\rm mD}(L_3)$ corresponding to the parameters
$\epsilon_1^{\rm o}$, $\epsilon_{\rm 1}^{\rm tD}$ and $\epsilon_{\rm 1}^{\rm mD}$ respectively.
The parameter choices in Figures \ref{fig:matfig503} through \ref{fig:matfig504b}, are $L_3=0.1$ (prolate spheroid),
$L_3=1/3$ (sphere) and $L_3=0.499$ (oblate spheroid) which is close to the limiting case $L_3=1/2$ expressed in \eqref{eq:Ljcond}.

From these examples, it is seen how the increased conductivity and losses (Figures \ref{fig:matfig503b} and \ref{fig:matfig504b}) limits
the usefulness of the local heating. But even in the latter example, where $\sigma=1$\unit{S/m}, the potential of local heating
amounts to a relative absorption ratio of about 10:1. In the case with the mismatched Drude model, it is interesting to see how a
prolongation of the spheroid lowers the resonance frequency, and a flattening of the spheroid yields a higher resonance frequency.


\section{Summary}
A new general formula has been derived for the absorption cross section of small ellipsoidal particles that are surrounded by lossy media.
The new formula is expressed explicitly in terms of the polarizability of the particle and can be used to define 
an optimal plasmonic resonance for a given surrounding medium. 
The new formula can be derived from general Mie scattering theory for a spherical
particle in a lossy medium which generalizes to particles of ellipsoidal shape in the limiting case with small particles.
The formula can furthermore be derived directly from the knowledge about the static solution to the ellipsoidal polarizability problem.
A canonical example is presented based on the polarizability of a homogeneous spheroid.
The example shows how an optimal plasmonic resonance can be designed based on a tuned Drude model 
and illustrates the typical shape dependent resonance frequency of the surface plasmon. The numerical example is
furthermore motivated by the medical application with radiotherapeutic hyperthermia based on electrophoretic heating of 
gold nanoparticle suspensions using microwave radiation.

\section*{Acknowledgment}
This work was supported by the Swedish Foundation for Strategic Research (SSF).



\begin{IEEEbiographynophoto}{Mariana~Dalarsson}
received the M.S.~degree in microelectronics in 2010, licentiate degree in electromagnetic theory in 2013 and Ph.D.~degree in 
electromagnetic theory in 2016 from the Royal Institute of Technology, Stockholm, Sweden. Since 2016, she is a postdoctoral 
researcher at the Department of Physics and Electrical Engineering, Linn\ae us University. 
Her research interests are in mathematical physics, metamaterials, electromagnetic wave propagation and absorption, inverse problems and imaging.
\end{IEEEbiographynophoto}

\begin{IEEEbiographynophoto}{Sven~Nordebo}
received the M.S.~degree in electrical engineering from the Royal Institute of Technology, Stockholm, Sweden, in 1989, 
and the Ph.D.~degree in signal processing from Lule{\aa} University of Technology, Lule{\aa}, Sweden, in 1995. 
He was appointed Docent in signal processing at Blekinge Institute of Technology, in 1999.
Since 2002 he is a Professor of Signal Processing at the 
Department of Physics and Electrical Engineering, Linn\ae us University. 
His research interests are in statistical signal processing, optimization, electromagnetics, wave propagation, inverse problems and imaging.
\end{IEEEbiographynophoto}

\begin{IEEEbiographynophoto}{Daniel~Sj\"{o}berg}
received the M.Sc.~degree in engineering physics and Ph.D.~degree in engineering, electromagnetic theory from Lund University, 
Lund, Sweden, in 1996 and 2001, respectively. In 2001, he joined the Electromagnetic Theory Group, Lund University, where, in 2005, 
he became a Docent in electromagnetic theory. He is currently a Professor and the Head of the Department of Electrical and Information Technology, 
Lund University. He serves as the Chair of Swedish URSI Commission B Fields and Waves since 2015. His research interests are in electromagnetic 
properties of materials, composite materials, homogenization, periodic structures, numerical methods, radar cross section, 
wave propagation in complex and nonlinear media, and inverse scattering problems.
\end{IEEEbiographynophoto}

\begin{IEEEbiographynophoto}{Richard~Bayford}
received the M.Sc.~degree in engineering from Cranfield Institute of Technology, UK, in 1981 and Ph.D. degree in engineering, from Middlesex University, UK, in 1994. 
He is currently Professor of Bio-Modelling, Head of Biophysics at the Middlesex University Centre for Investigative Oncology, 
Head of Biophysics and Engineering group, and Visiting Professor at UCL, Department of Electrical and Electronic Engineering, UK. 
His expertise is in biomedical imaging, bio-modelling, nanotechnology, deep brain stimulation, tele-medical systems, instrumentation and biosensors. 
He has had long collaborations with research groups both in the UK and overseas. 
He has published over 270 scientific papers and served as Editor-in-Chief for Physiological Measurements.
\end{IEEEbiographynophoto}

\end{document}